\documentclass[9pt,twocolumn,twoside]{osajnl}

\journal{none} 

\setboolean{shortarticle}{false}

\title{Quasi-static Optical Parametric Amplification}

\author[1,2,*]{Marc Jankowski}
\author[1,*]{Nayara Jornod}
\author[1]{Carsten Langrock}
\author[3]{Boris Desiatov}
\author[4]{Alireza Marandi}
\author[3]{Marko Lon\v{c}ar}
\author[1]{Martin M. Fejer}

\affil[1]{Edward L. Ginzton Laboratory, Stanford University, Stanford, California 94305, USA}
\affil[2]{NTT Research Inc. Physics and Informatics Labs, 940 Stewart Drive, Sunnyvale, California}
\affil[3]{John A. Paulson School of Engineering and Applied Sciences, Harvard University, Cambridge, Massachusetts 02138, USA}
\affil[4]{Department of Electrical Engineering, California Institute of Technology, Pasadena, California 91125, USA}

\affil[*]{These authors contributed equally to this work: marc.jankowski@ntt-research.com, nayaraj@stanford.edu}

\begin{abstract}

High-gain optical parametric amplification is a crucial nonlinear process used both as a source of coherent infrared light and as a source of non-classical light. In this work, we experimentally demonstrate an approach to optical parametric amplification that enables extremely large parametric gains with low energy requirements. In conventional nonlinear media driven by femtosecond pulses, multiple dispersion orders limit the effective interaction length available for parametric amplification. Here, we use the dispersion engineering available in periodically poled thin-film lithium niobate nanowaveguides to eliminate several dispersion orders at once. The result is a quasi-static process; the large peak intensity associated with a short pump pulse can provide gain to signal photons without undergoing pulse distortion or temporal walk-off. We characterize the parametric gain available in these waveguides using optical parametric generation, where vacuum fluctuations are amplified to macroscopic intensities. In the unsaturated regime, we observe parametric gains as large as 71 dB (118 dB/cm) spanning 1700 - 2700 nm with pump energies of only 4 picojoules. When driven with pulse energies >10 pJ, we observe saturated parametric gains as large as 88 dB (>146 dB/cm). The devices shown here achieve saturated optical parametric generation with orders of magnitude less pulse energy than previous techniques.

\end{abstract}

\setboolean{displaycopyright}{true}

\begin{document}

\maketitle

\section{Introduction}\label{introduction}

Optical parametric amplification (OPA) in quadratic ($\chi^{(2)}$) nonlinear media is an indispensable process in both quantum and classical nonlinear optics. When seeded by a coherent signal or embedded in a resonator to form an optical parametric oscillator, OPA can be used to produce both broadband and broadly tunable coherent light at otherwise difficult to access wavelengths, and is an essential tool in spectroscopy~\cite{Schliesser2012}. An advantage of parametric sources when compared to laser gain media is that the gain bandwidth available for OPA is determined by the group velocity dispersion (GVD) of the nonlinear medium rather than the linewidth of a resonant transition; OPA can produce light at any wavelength within the transparency window of the nonlinear crystal, and octave-spanning gain bandwidths becomes possible when the signal and idler are generated around to the zero GVD wavelength of the nonlinear medium~\cite{Lin2020,Kuo:06,Crouch1988}. In the absence of a coherent seed the pump field provides gain to vacuum fluctuations at the phase-matched signal and idler wavelengths, and can be used as a source of biphotons~\cite{URen2005GenerationOP,Mosley2008,Harris2007}, squeezed light~\cite{Crouch1988, Kashiwazaki2020}, and highly non-classical states of light~\cite{Reid1993, onodera2019nonlinear}. In almost every application both high gain and low energy operation are desirable. In practice, the $\chi^{(2)}$ nonlinearities used to drive OPA are inherently weak and therefore require large pump intensities and long interaction lengths to achieve large parametric gains. As a result, state-of-the-art high-gain OPAs based on diffused waveguides operate with nanojoules of pump pulse energy~\cite{Xie:04}.

This work builds on recent progress in dispersion-engineered periodically poled thin-film lithium niobate (TFLN) nanophotonics to realize large parametric gains using only picojoules of pump pulse energy. These results are made possible by two features unique to the TFLN platform: wavelength-scale modal confinement and quasi-static operation. The tight spatial confinement found in TFLN devices increases the field intensity 20-fold relative to diffused waveguides. Early demonstrations of continuous-wave- (CW-) pumped second-harmonic generation (SHG) demonstrated a corresponding 20-fold reduction of the power required for efficient operation~\cite{Wang:18,Rao2019,Zhao2020}. Quasi-static operation occurs when the group velocity mismatch (GVM) and GVD of the interacting waves are negligible. In this limit, nonlinear processes can be driven with femtosecond pulses without any of the interacting harmonics experiencing significant temporal walk-off or chirp. Quasi-static devices enable low energy operation by combining long interaction lengths with the large field intensities associated with femtosecond pulses. Recent work has shown that quasi-static operation can be achieved by using the geometric dispersion present in TFLN waveguides to simultaneously eliminate both GVM and GVD, and the resulting devices demonstrated saturated SHG with less than 100 femtojoules of pulse energy~\cite{Jankowski:20}. 


\begin{figure*}[t!]
\includegraphics[width=\linewidth]{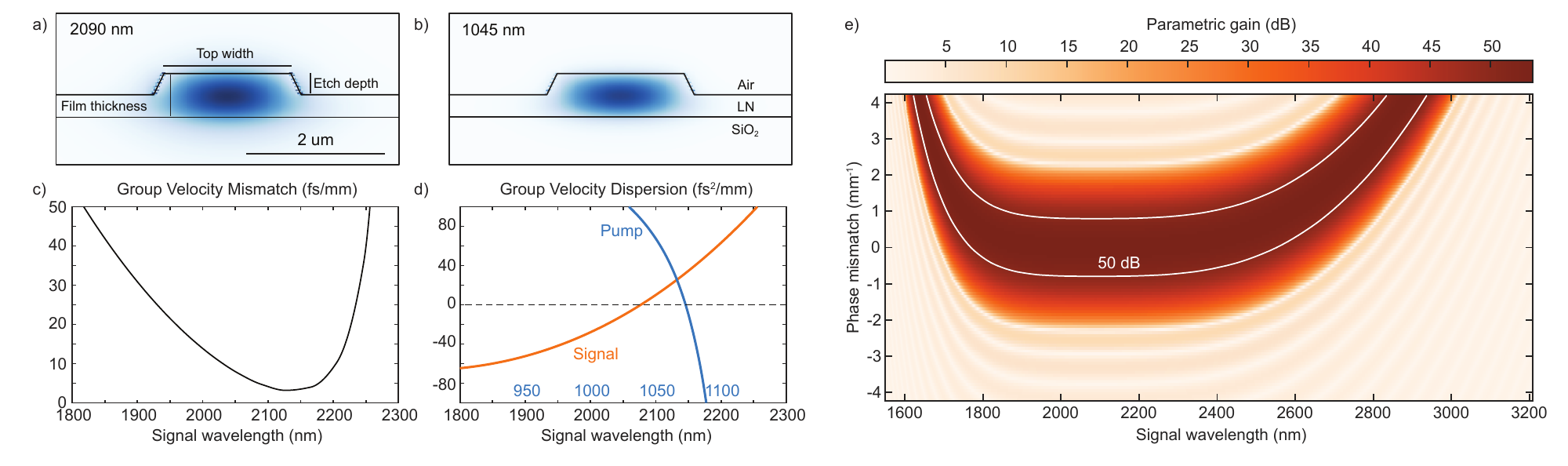}
\caption{\textbf{a,b}, Waveguide cross-section and normalized electric field of the simulated TE$_{00}$ mode, a) signal and b) pump. \textbf{c}, Simulated group velocity mismatch ($\Delta k'$), and \textbf{d}, group velocity dispersion (orange line: $k''_{\omega}$, blue line: $k''_{2\omega}$) as a function of the signal wavelength. The wavelengths associated with $k''_{2\omega}$ are shown in blue. \textbf{e}, Simulated OPG bandwidth for a pump pulse energy of 1 pJ. Solid white contour lines correspond to 50 dB of gain.}
\label{fig:fig1}
\end{figure*}

Here we report one of the first demonstrations of high-gain OPA in a quasi-static nonlinear photonic device. This work proceeds in four sections. Section \ref{method} discusses the theory and design of quasi-static OPAs, and establishes the key dispersion orders that must be suppressed in order to achieve quasi-static operation. In Section \ref{results} we experimentally characterize the parametric gain and bandwidth of a quasi-static OPA by using optical parametric generation (OPG). When driven by 80-fs pump pulses, we observe unsaturated gains as large as 71 dB (118 dB/cm) across a micron of bandwidth (43\% fractional bandwidth) for pump-pulse energies of 4 picojoules. For pump-pulse energies in excess of 10 pJ we observe saturated parametric gains as large as 88 dB, which provides a lower bound for the power gain coefficient of 146 dB/cm. In the saturated regime the generated signal clamps at 15\% of the input pump pulse energy, and both the generated signal and depleted pump undergo spectral broadening to form a supercontinuum spanning 900-2900 nm. Section \ref{Discussion} analyzes the observed behavior of quasi-static OPAs using numerical simulations and establishes the mechanisms responsible for both the power clamping and the spectral broadening. Section \ref{Conclusion} concludes this work by putting these results in context. The devices shown here achieve saturated optical parametric generation with two orders of magnitude less pulse energy than previous state-of-the-art devices based on diffused waveguides.

\section{Theory and Design}\label{method}

Throughout this article, we restrict our focus to the case of quasi-phasematched (QPM) degenerate OPA in a waveguide, where pump pulses centered around frequency $2\omega$ generate a signal around $\omega$. The evolution of the time-domain pump and signal envelopes in a frame moving with the group velocity of the signal are given by the coupled-wave equations (CWEs) 
\begin{subequations}
\begin{align}
\partial_z A_\omega(z,t) &= -i\kappa A_{2\omega}(z,t)A_\omega^*(z,t) \exp(-i\Delta k z)\label{eqn:CWE11}\\
&+ \hat{D}_\omega A_\omega(z,t),\nonumber\\
\partial_z A_{2\omega}(z,t) &= -i\kappa A_\omega^2(z,t) \exp(i\Delta k z)\label{eqn:CWE12}\\
&-\Delta k' \partial_t A_{2\omega}(z,t)+\hat{D}_{2\omega} A_{2\omega}(z,t),\nonumber
\end{align}
\end{subequations}
where $A_\omega$ is the complex field amplitude of the signal generated around frequency $\omega$, normalized such that $P_\omega=|A_\omega|^2$ corresponds to the instantaneous power of the signal centered around $\omega$. $\Delta k'=k'_{2\omega}-k'_{\omega}$ is the group velocity mismatch between the pump and the generated signal. The dispersion operator $\hat{D}_\omega=\sum_{j=2}^{\infty} \left((-i)^{j+1}k_\omega^{(j)}/j!\right)\partial_t^j$ contains contributions from second- and higher- order dispersion, where $k_\omega^{(j)}$ represents the $j^{th}$ derivative of the propagation constant $k$ at frequency $\omega$. We note here that for frequency derivatives of orders three or less, the usual prime notation will be used interchangeably with the notation defined above, $k_\omega^{(3)}=k_\omega'''$. The nonlinear coupling is given by
\begin{equation}
\kappa = \frac{\sqrt{2 Z_0} \omega d_\mathrm{eff}}{c n_\omega \sqrt{n_{2\omega}A_{\mathrm{eff}}}},\label{eqn:kappa}
\end{equation}
where $Z_0$ is the impedance of free space, $n_\omega$ is the effective index of the relevant mode at frequency $\omega$, $d_\mathrm{eff} = 2d_{33}/\pi$ is the effective nonlinear coefficient, and $A_\mathrm{eff}$ is the effective area of the nonlinear interaction as determined by the overlap integral of the interacting modes~\cite{Jankowski:21}.

In the absence of phase-mismatch ($\Delta k = 0$), GVM ($\Delta k' = 0$), higher order dispersion ($\hat{D}=0$), and pump depletion, the evolution of the signal pulse is given by $A_\omega(z,t) = A_\omega(0,t)\exp(\gamma(t)z)$ where $\gamma(t) = \kappa A_{2\omega}(0,t)$ is the unsaturated gain coefficient. As expected, in the limit where dispersion is negligible to arbitrary order the portion of the signal located around the peak of the pump ($t=0$) experiences a large parametric gain $\gamma_\mathrm{pk}=\kappa A_{2\omega}(0,0)$ due to the large local field intensity. In the presence of GVM, Eqns. \ref{eqn:CWE11}-\ref{eqn:CWE12} can be integrated to find that the gain experienced by a signal pulse around $t=0$ is given by $\exp(\gamma_\mathrm{pk} 2\tau/\Delta k')$, where we have assumed $A_{2\omega}(0,t)=\sqrt{U}\mathrm{sech}\left[(t+\Delta k'L/2)/\tau\right]/\sqrt{2\tau}$. Here, $U$ is the input pulse energy to the waveguide and $L$ is the length of the nonlinear crystal. In this case, the peak of the pump pulse walks with respect to the signal from $t=-\Delta k' L/2$ to $t=\Delta k' L/2$ during propagation and the effective interaction length for the given portion of the signal around $t=0$ is reduced to
\begin{equation}
L_\mathrm{eff}=2\tau/\Delta k'.\label{eqn:interactionlength}
\end{equation}
Eqn. \ref{eqn:interactionlength} illustrates the role dispersion plays in suppressing parametric gain. Reducing the GVM increases the effective interaction length for a given input pulse duration $\tau$, thereby realizing an exponential increase in the generated signal power.

In practice, several dispersion orders restrict both the pump bandwidth available for OPA and the generated signal bandwidth. A more complete description of OPA that accounts for dispersion to arbitrary order can be obtained by solving the CWEs in the frequency domain (see supplemental document). The pump bandwidth available to contribute parametric gain is limited to leading order by the temporal walkoff, which becomes negligible when $\Delta k' L/\tau \ll 1$. Higher order contributions, such as chirp and third order dispersion are negligible when $k_{2\omega}^{(n)}L/(\tau^n n!)\ll 1$. Typical values for the pump pulse duration and dispersion terms are given by $\tau = 100/1.76$ fs, $\Delta k' = 100$ fs/mm, $k_{2\omega}''=100$ fs$^2$/mm, and $k_{2\omega}'''=1000$ fs$^3$/mm; dispersion orders beyond the temporal walkoff typically only restrict the pump bandwidth for waveguides approaching lengths of 10 centimeters. When both temporal walk-off and higher dispersion orders of the pump are negligible, the field gain coefficient of a signal and idler detuned $\pm \Omega'$ from degeneracy can be approximated as
\begin{subequations}
\begin{align}
&\gamma(\Omega') = \sqrt{\gamma_\mathrm{pk}^2-\left(\Delta k_\omega(\Omega')/2\right)^2},\\
\Delta k_\omega(\Omega') = k_{2\omega}&(2\omega) - k_{\omega}(\omega + \Omega') - k_{\omega}(\omega - \Omega') -2\pi/\Lambda_G,\nonumber
\end{align}
\end{subequations}
where $\Lambda_G$ is the period of the QPM grating. Exponential growth occurs for wavelengths around $\Delta k_\omega(\Omega') = 0$, with oscillatory solutions occurring for wavelengths where $\Delta k_\omega(\Omega') > 2\gamma_\mathrm{pk}$. The series expansion of $\Delta k_\omega(\Omega')$ with respect to $\Omega'$ only contains even dispersion orders. Therefore, the generated OPA bandwidth is limited to leading order by the GVD per gain length of the fundamental, $k_\omega''/\gamma_\mathrm{pk}$. In the absence of GVD, the generated signal bandwidth is dominated by the fourth-order dispersion per gain length, $k_\omega^{(4)}/\gamma_\mathrm{pk}$. We therefore prioritize simultaneous reductions of $\Delta k'$ and $k_\omega''$ to enable both long interaction lengths and broad gain bandwidths.

For OPG, the photon flux generated at $\Omega'$ is determined by 
\begin{equation}
\Phi(\Omega') \propto  f_R (\gamma_\mathrm{pk}z)^2 \left|\frac{\mathrm{sinh}(\gamma(\Omega')z)}{\gamma(\Omega')z}\right|^2,
\end{equation}
where $f_R$ is the repetition frequency of the pump laser. Therefore, OPG provides a direct measurement of both the gain and the bandwidth associated with OPA. For large peak gain coefficients, $\gamma_\mathrm{pk}$, the parametric gain can be determined by measuring the scaling of the power generated by OPG with respect to input pump power $P_\mathrm{OPG}\propto \exp(2\kappa \sqrt{P_\mathrm{in,pk}} L)$. Similarly, the OPA bandwidth can be determined by the wavelengths that satisfy $2\gamma_\mathrm{pk} = \Delta k_\omega (\Omega')$. At these wavelengths, $\gamma(\Omega')$ becomes imaginary and $\Phi(\Omega') \propto  f_R (\gamma_\mathrm{pk}z)^2 \mathrm{sinc}^2(|\gamma(\Omega')|z)$. For wavelengths slightly past this point $\mathrm{sinc}^2(|\gamma(\Omega')|z)=0$ and the photon flux generated by OPG vanishes.


Figure \ref{fig:fig1} summarizes the relevant design parameters of a nanophotonic waveguide designed for broadband OPA driven by short pulses centered around 1045 nm. We use 6-mm-long waveguides fabricated in a 700-nm thin film of 5\% MgO-doped x-cut lithium niobate (NANOLN). These waveguides have a top width of 1850~nm and an etch depth of 340~nm (Fig.~\ref{fig:fig1}a). The resulting $\Delta k'$ and $k_\omega''$ are shown in Fig.~\ref{fig:fig1}c and d, respectively, showing $\Delta k'\sim~5$~fs/mm,  $k_\omega''\sim$~6~fs$^2$/mm, and $k_{2\omega}''\sim$~80 fs$^2$/mm. For the propagation lengths considered here and the 80-fs pulse duration of the pump, both temporal walk-off and pump dispersion are negligible, and the entire pump bandwidth may contribute to OPG. Figure \ref{fig:fig1}e shows the calculated gain available for OPA as a function of the phase-mismatch between the pump and signal, $\Delta k_\omega(\Omega'=0)$, and the wavelength of the generated signal. Here we assume a pump pulse energy of 1 pJ, and a pulse duration of 80 fs. When the fundamental and second harmonic are close to phase-matching, the calculated gain spectrum exceeds 50~dB across nearly a micron of bandwidth ($\sim$1700 - 2600~nm). We note here that in the limit of large peak gain a slight offset of the phase-mismatch ($\Delta k_\omega(\Omega'=0)=$ 1-2 mm$^{-1}$) has a negligible effect on the observed parametric gain and can be used to enhance the gain bandwidth by hundreds of nanometers~\cite{Crouch1988}. The waveguides studied in the experiment have a poling period of 5.12 $\mu$m to achieve broadband gain centered around 2090 nm.


\begin{figure*}[ht!]
\centering
\includegraphics[width=\linewidth]{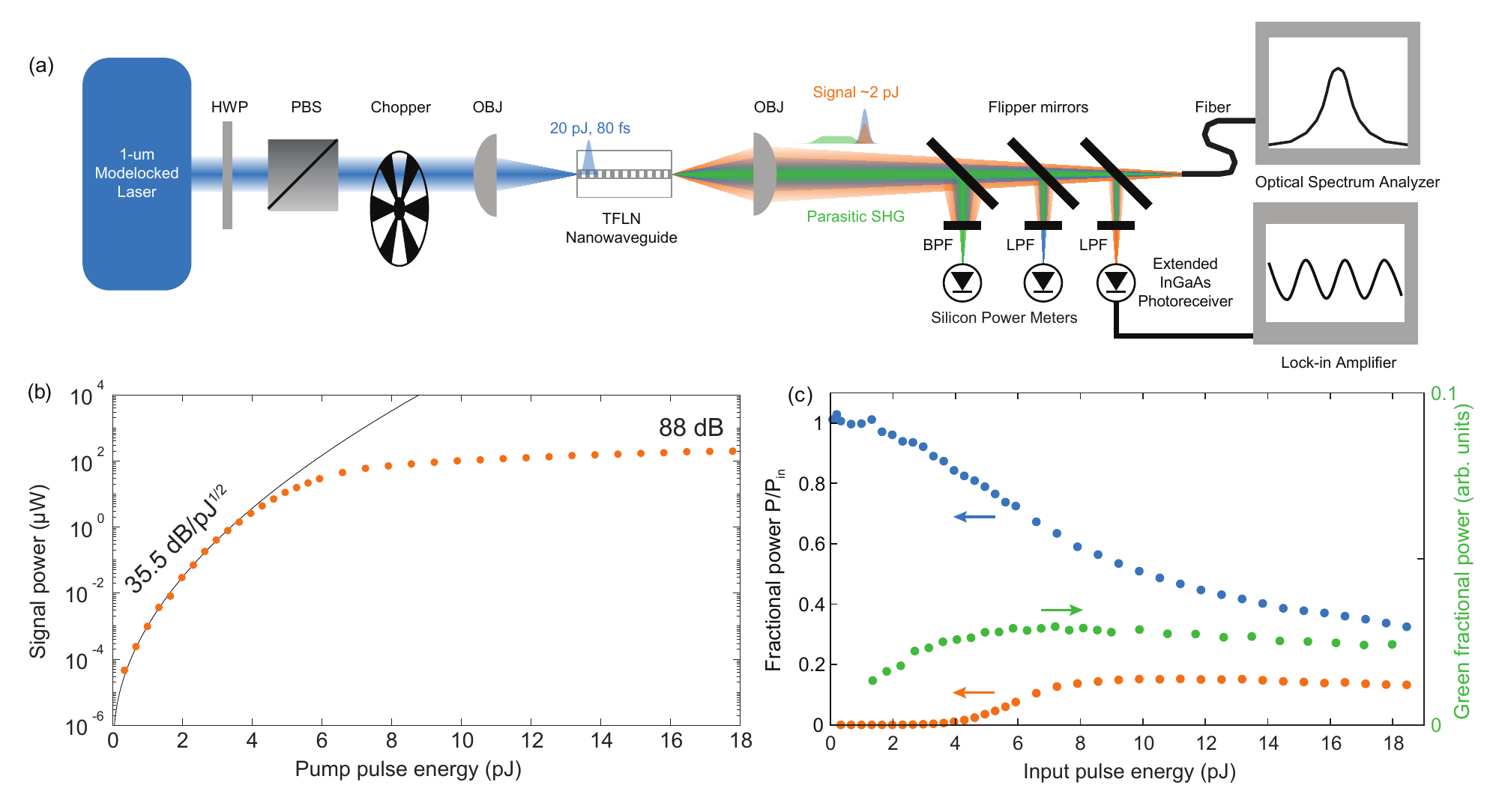}
\caption{\textbf{a}, Experimental setup. HWP, half-wave plate; PBS, polarizing beamsplitter; OBJ, reflective objective lens; LPF, long pass filter; BPF, band pass filter. \textbf{b}, Semi-logarithmic scale of the generated signal power, showing a gain of 35.5 dB/$\sqrt{\mathrm{pJ}}$ (solid black line) and a saturated gain of 88 dB. \textbf{c}, Left axis: OPG efficiency (orange), and pump depletion (blue) as a function of input pulse energy; Right axis: Parasitic green conversion efficiency, in arbitrary units.  }
\label{fig:fig2}
\end{figure*}

\section{Results}\label{results}
\subsection{\label{Experimental setup}Experimental setup}
We characterize OPG in these TFLN nanowaveguides using the experimental setup shown in Fig.~\ref{fig:fig2}a, and note that phase-sensitive amplification of short pulses is reported for similar devices in~\cite{Ledezma:21}. 80-fs pulses centered around 1045~nm from a Yb:fiber laser (Menlo Systems Orange A, repetition frequency 100~MHz) are coupled into the waveguides using a reflective objective lens (Thorlabs LMM-40X-P01) and a similar objective is placed at the output to collect the generated signal and transmitted pump for characterization. We note here that the diffraction-limited spot produced by focusing a Gaussian beam through a reflective objective exhibits a poor overlap with the waveguide mode and therefore limits the coupling efficiency to be on the order of a few percent. The results presented here correspond to a measured input coupling efficiency of 2.2\% and an estimated  collection efficiency from the waveguide of 30\%. While this input coupling efficiency is small relative to conventional high-NA objective lenses, this approach eliminates both chromatic aberration and pre-chirping of the pump pulses. These features greatly simplify the analysis of the depleted pump and generated signal collected from the chip, and in all further discussions we report the pulse energies and conversion efficiencies inside the nonlinear waveguide. The coupling and collection efficiencies for future generations of devices could be improved to $>$90\% by using tapered edge couplers to expand the waveguide modes at the input and output of the waveguide~\cite{Hu:21}.


We monitor three powers at the output of the waveguide: the generated signal, the pump, and parasitic green light generated by SHG of the pump. The generated signal is measured using an extended InGaAs photoreceiver (NewFocus Model 2034) with a wavelength sensitivity range of 800 – 2200~nm and a lock-in amplifier (SRS SR830). An additional longpass filter (Thorlabs FELH1350) is placed before the photoreceiver to remove any contribution of the transmitted pump from reaching the detector. The transmitted pump and parasitic green are characterized with a silicon photodiode (Thorlabs S130C). We filter the light output from the chip using a longpass filter (Thorlabs FELH900) to measure the pump and a bandpass filter (Thorlabs FGS900-A) for the green light.
We characterize the spectrum of the pump and signal by collecting the light output from the waveguide into a zirconium fluoride fiber (Thorlabs P1-23Z-FC-5) and sending the signal to an optical spectrum analyzer. The downconverted signal is detected using a Yokagawa AQ6376 with a span of 1500-3400~nm, and wavelengths around the pump are detected using a Yokagawa AQ6370C with span of 800-1500~nm.

\subsection{\label{gain}Gain characterization}
We observe the onset of OPG with only 60~fJ of in-coupled pulse energy, limited by the noise floor of our photoreceiver and the 1-s integration time used for lock-in detection. 
At low power we observe the expected $\sqrt{P_\mathrm{in}}$ dependence of the exponential growth on the logarithmic scale (Fig.~\ref{fig:fig2}b), and find a fitted gain coefficient of 35.5 dB/$\sqrt{\mathrm{pJ}}$ (black line), in reasonable agreement with the 55 dB/$\sqrt{\mathrm{pJ}}$ predicted by quasi-static theory. We note here that previous characterization of these waveguides using SHG found normalized efficiencies ($\kappa^2$) consistent with theory~\cite{Jankowski:20}. Therefore, the observed deviations between theory and experiment may be due to either input coupling of the pump to higher order spatial modes of the waveguide, which would reduce the amount of pump power in the TE$_{00}$ mode available for OPA, or depletion of the pump due to parasitic SHG. In Section \ref{Discussion}, we determine that input coupling of the pump to higher order spatial modes of the waveguide, rather than parasitic loss, is responsible for this reduction of the observed nonlinearity.

We note here that given the exponential scaling of the generated signal power with the incoupled pump power, the generated signal is extremely sensitive to misalignments between the waveguide and the incident pump. We observe drifts in the measured signal power of 1-3 dB that we attribute to slow temperature drifts of $\pm$0.1~C of the crystal mount, which misaligns the waveguide relative to the pump beam. To the greatest extent possible, the output power and generated spectra presented here are taken at the peak signal output power of these drifts. In principle, this alignment sensitivity may be greatly reduced using the techniques discussed above to expand the waveguide mode~\cite{Hu:21}.

For in-coupled pulse energies below 4 pJ, the signal experiences unsaturated gain. At a pump pulse energy of 4 pJ we find an unsaturated gain of 71 dB, corresponding to a power gain coefficient of 118 dB/cm. For in-coupled pulse energies larger than 4 pJ, the generated signal power saturates and becomes linear in the input pump energy. For a pump energy of 18~pJ we measured a signal power of 200~$\mu$W (2~pJ), which corresponds to a saturated gain of 88 dB (> 146 dB/cm). Here we calculate an effective saturated gain by comparing the power of the generated signal to the power measured in the unsaturated regime, e.g. we define the signal power of 200~$\mu$W at 18 pJ of pump energy as representing an additional 17 dB of gain relative to the signal power of 3.7~$\mu$W generated by a pump pulse energy of 4 pJ. We note, however, that in the saturated regime the signal generated by OPG undergoes spectral broadening and the power spectral density within the OPA bandwidth becomes a weak function of input pump power. Instead, most of the additional photons generated in the saturated regime occur at wavelengths outside of the OPA bandwidth.

\subsection{\label{efficiency}Conversion efficiency}
The conversion efficiency from pump to signal and the depletion of the input pump as a function of the input pulse energy are presented in Fig.~\ref{fig:fig2}c. While not pertinent to the goal of characterizing the gain and bandwidth of the OPA, through modeling of this saturated regime we can identify parasitic effects that must be considered in applications of these OPA devices. After the exponential growth at low powers, the conversion efficiency reaches a plateau of 15\% for pump pulse energies above 10~pJ.
The pump experiences depletion early on in the process, exceeding 10\% already for pulse energies larger than 4 pJ.
At higher pulse energy, the observed pump depletion ($>$60$\%$) is drastically stronger than the measured fractional power of the generated signal (15$\%$). We attribute this to parasitic SHG of the pump light. For all levels of pumping we observe green light scattering out of the waveguide, with a small fraction collected by the output objective. This scattering is due to a combination of surface roughness and lateral leakage into slab modes, the latter of which becomes more pronounced at short wavelengths~\cite{Boes:19}.

The right axis of Fig.~\ref{fig:fig2}c shows the collected green power due to parasitic SHG measured at the output of the waveguide. We note here that these curves \emph{do not} sum to one, and that the generated green can only be reported in arbitrary units. The green generated during propagation is a nonlinear function of the position in the waveguide. While low power devices generate green light along the full propagation length of the waveguide, strongly driven devices generate more green light near the input, before the pump is depleted, and therefore a smaller fraction of the generated green light reaches the output due to propagation loss. To better quantify the generated green, we repeated these measurements for both a narrower waveguide and an identical waveguide with a larger phase-mismatch. Neither of these waveguides produced measurable 2-$\mu$m signals, and in both cases the observed pump depletion was comparable to the pump depletion reported here. There are two possible explanations for the role played by the parasitic SHG. If the green light is generated by SHG of the input TE00 mode of the pump, then the pump depletion due to SHG may clamp the generated signal power. Conversely, if the green is generated by SHG of a higher order spatial mode of the pump, then the pump depletion due to SHG has no deleterious effect on OPA. In this case, the SHG signal becomes clamped at 10\% since only a fraction of the input pump was launched into the TE00 mode, which limits the amount of pump power available to deplete by OPA. We discuss these possibilities in more detail in Section \ref{Discussion}.


\subsection{Generated signal bandwidth}\label{bandwidth}

\begin{figure}[t]
\centering
\includegraphics[width=\linewidth]{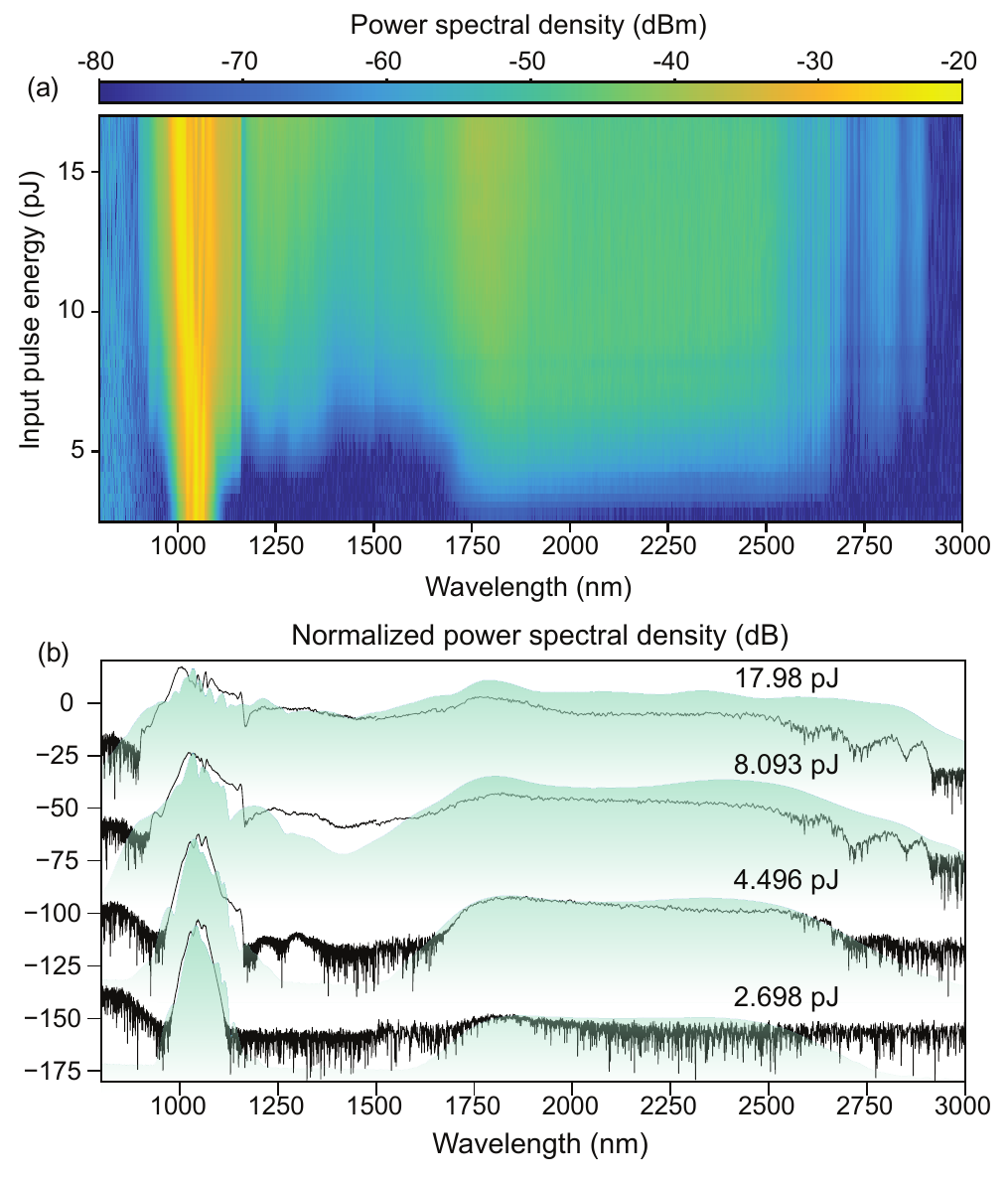}
\caption{\textbf{a}, Generated spectrum as a function of input pulse energy. \textbf{b}, Selection of measured power spectral density compared with simulations. Adjacent traces are shifted by 40~dB to facilitate readability. Light green: simulation.}
\label{fig:fig3}
\end{figure}

The experimentally measured OPG spectrum is presented in Fig.~\ref{fig:fig3}a. At input energies below 4~pJ, corresponding to the unsaturated regime, we observe the formation of a broadband signal spanning from 1700~nm to 2700~nm (70~THz), in good agreement with the expected OPG bandwidth. The generated signal spans >60 THz at the -10 dB level, corresponding to a fractional bandwidth >43\%. We note, however, that even for low input powers the pump exhibits some spectral broadening. Similar amounts of pump spectral broadening are present in waveguides that are not phase-matched for OPG, which suggests this pump broadening is due to either self-phase modulation or coupling between the pump and the parasitic SHG. When the in-coupled pump energy is increased beyond 4~pJ, the signal partially depletes the pump and both are observed to undergo rapid spectral broadening. For pulse energies in excess of 7~pJ, the pump and signal merge into a broadband spectrum. The dips in the spectrum around 2750~nm and 2850~nm are due to OH-absorption in the silica substrate and lithium niobate, respectively, and can be reduced by annealing the bulk materials before fabricating the thin films~\cite{Schwesyg:10}. Any light generated beyond 2900 nm is absorbed in the silica and not observed here. Alternative substrates, such as sapphire, can evade this limitation and allow operation at wavelengths as long as 4.5 $\mu$m, where multiphonon absorption in LN becomes significant~\cite{Mishra:21}. We close this section by noting that while this generated supercontinuum is \emph{not} coherent since it was seeded by vacuum fluctuations, it may represent a novel approach to spectral broadening. Similar effects were observed for saturated OPG in~\cite{Kuo:06}, but to the best of our knowledge this process has not been studied in detail.


\section{Discussion}\label{Discussion}

While the gain and bandwidth generated by OPG at low power exhibit reasonable agreement with theory, a number of unexpected behaviors were observed in these waveguides. These effects include a reduction of the observed parametric gain to 35.5 dB/$\sqrt{\mathrm{pJ}}$, the generation of parasitic green light, clamping of the generated signal power to 15\% of the input pump power, and spectral broadening of the pump and signal in the depleted regime. To better understand the interplay of each of these mechanisms, we simulate the evolution of the interacting waves in the presence of both $\chi^{(2)}$ and third-order ($\chi^{(3)}$) nonlinearities using an adaptive split-step Fourier method that includes three envelopes in the equations of motion: the signal around 2~$\mu$m, seeded by semi-classical vacuum fluctuations, the pump around 1~$\mu$m, and parasitic green light around 520~nm formed by SHG of the pump. The multi-envelope approach used here allows us to develop intuition about the various interacting nonlinear processes by isolating each of the relevant nonlinear interactions~\cite{Phillips:11}.

Figure~\ref{fig:fig3}b shows a comparison of the measured (black) and simulated (light green) spectra for a selection of pump pulse energies, showing qualitative agreement between simulation and experiment. To isolate the role of each nonlinear interaction, we repeat these simulations in the absence of the nonlinear polarization generated by the interaction in question. We note here that repeating these simulations with and without $\chi^{(3)}$ nonlinearities, including both self- and cross-phase modulation, yields nearly identical evolution for the generated harmonics, and therefore the observed spectral broadening of the pump and signal must be dominated by quadratic nonlinearities. We neglect $\chi^{(3)}$ nonlinearities in any further discussion.

We first consider the case of OPG in the absence of parasitic SHG and any $\chi^{(3)}$ processes. In this case, the evolution of the signal in the absence of pump depletion is identical to the simple analytic model described above. In the saturated regime, the generated signal can be up to 40\% of the input pump energy, and both the pump and signal undergo spectral broadening with resulting bandwidths comparable to those observed in experiment. This suggests that the spectral broadening that occurs for large pump-pulse energies is due to pump depletion. We note here that to the best of our knowledge, a heuristic model for how pump depletion causes spectral broadening is absent from the OPG literature. Further study of these broadening mechanisms in the presence of a coherent seed and the development of reduced models that can explain this effect will be the subject of future work.

We next examine the clamping of the generated signal power in saturation. To determine the amount of parasitic SHG by the TE00 mode needed to explain the conversion efficiency observed in experiment, we take both the nonlinear coupling and phase-mismatch between the pump and generated green light as fitting parameters. We find that the depletion of the pump by parasitic SHG can only explain the observed conversion efficiencies for extremely large amounts of nonlinear coupling. When the fitted $\kappa$ and $\Delta k$ are compared to values determined by mode simulations, we find that $\kappa^2$ must be increased by 50\%, and $\Delta k$ must be decreased by an order of magnitude. Furthermore, we note that no amount of parasitic SHG can simultaneously account for both the reduction of parametric gain at low pump energies and the observed conversion efficiency for large parametric gains. This suggests that the input pump is coupled into multiple spatial modes. In this case, the observed gain for small pump powers can be explained by a coupling efficiency of 41\% of the pump power into the TE00 spatial mode. This also clamps the generated signal power at 16\% of the total input power, since only a fraction of the input pump power is available in the TE00 mode. The remaining power coupled into higher-order spatial modes of the pump undergo parasitic SHG and scatter out of the waveguide due to lateral leakage. The leading contribution to pump depletion is due to SHG of the TE20 mode of the pump into the TE20 mode of the green. This process is nearly phase-matched by the QPM period used for OPA (4.99 $\mu$m and 5.12 $\mu$m, respectively). Accounting for both OPG of the TE00 mode and SHG of the TE20 mode, we find a pump depletion of 55\% for an input pulse energy of 18 pJ, in reasonable agreement with experiment.

\begin{figure}[t]
\centering
\includegraphics[width=\linewidth]{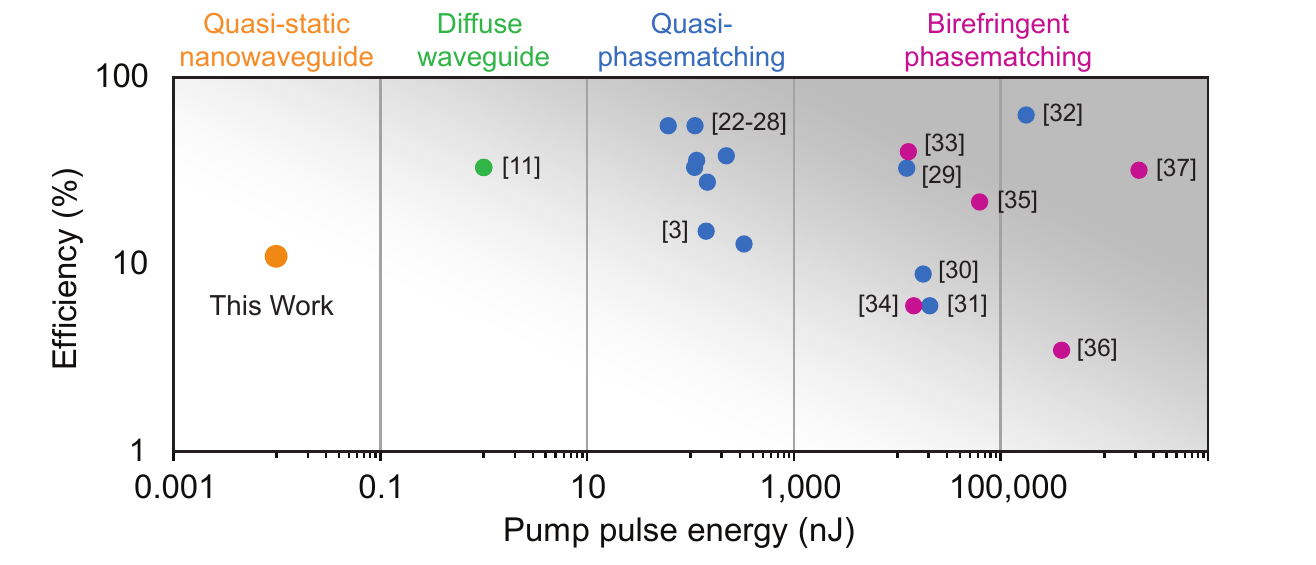}
\caption{Comparison of the energy required to achieve saturated OPG by quasi-static nonlinear photonic devices with previous demonstrations in bulk nonlinear media~\cite{Kuo:06, Aadhi:17, Marchese:2005, Linnenbank:14, Fan:17, Linnenbank:16, Sudmeyer:02, Galvanauska:1997, Xu:15, Prakash:08, Levenius:11, Zhao:06, DiTrapani:95, Piccoli:14, Chalus:10, Manzoni:2009, Kumar:12} and diffuse waveguides~\cite{Xie:04}.}
\label{fig:fig4}
\end{figure}

We close this section by placing these results in context. Even with the non-idealities observed here, the extremely high parametric gain coefficient of quasi-static OPA devices in comparison to previous approaches can be seen by considering the energy required to reach saturated OPG (Figure~\ref{fig:fig4}). Historically, the most common route toward high gain OPAs has been QPM interactions in bulk media (Figure~\ref{fig:fig4}, blue circles), with typical devices achieving saturated operation with 100's of nJ of pulse energy. State-of-the-art devices based on diffuse waveguides have achieved saturation using 1 nJ (Figure~\ref{fig:fig4}, green circle). A conservative estimate for the energy requirements of the devices reported here suggests we have achieved saturated OPG with 10 pJ of pump pulse energy. Restricting our focus to interactions between TE00 modes suggests this number may be as low as 5 pJ of pump pulse energy. The inefficient coupling to the TE00 mode that limited the performance of this device could be addressed in future designs by using tapers to both eliminate higher order spatial modes and expand the TE00 mode~\cite{Hu:21}.


\section{Conclusion}\label{Conclusion}

In summary, we have experimentally demonstrated high gain OPA and saturated OPG with only picojoules of pulse energy. This low energy operation was made possible by a new generation of dispersion-engineered $\chi^{(2)}$ nonlinear nanophotonic waveguides. The nanowaveguides used here achieve wavelength-scale transverse confinement of the interacting modes and use dispersion engineering to simultaneously eliminate temporal walk-off and group velocity dispersion. Both the resulting field intensities and effective interaction lengths for femtosecond pulses exceed previous approaches by more than an order of magnitude, thereby enabling large parametric gains with orders of magnitude less pulse energy than previous state-of-the-art technologies. We believe that dispersion-engineered quasi-phasematched nanowaveguides such as those shown here represent a crucial step towards chip-scale parametric devices that operate with exceptionally low power, including sources of optical frequency combs and nonclassical light.


\begin{backmatter}
\bmsection{Funding} The authors acknowledge support from National Science Foundation (NSF) (ECCS-1609688, EFMA-1741651, CCF-1918549); Department of Energy (DoE) (DE-AC02-76SF00515); Army Research Laboratory (ARL) (W911NF-15-2-0060, 48635-Z8401006); Army Research Office (ARO) (W911NF-18-1-0285), and the Swiss National Science Foundation (SNSF-P400P2-194369).

\bmsection{Acknowledgments} The authors wish to thank NTT Research for their financial and technical support. Electrode patterning and poling was performed at the Stanford Nanofabrication Facility, the Stanford Nano Shared Facilities (NSF award ECCS-2026822), and the Cell Sciences Imaging Facility (NCRR award S10RR02557401). Patterning and dry etching was performed at the Harvard University Center for Nanoscale Systems (CNS), a member of the National Nanotechnology Coordinated Infrastructure (NNCI) supported by the National Science Foundation. Facet preparation was done by Disco Hi-Tec America, Inc. The authors thank Edwin Ng for helpful discussions.

\bmsection{Disclosures} MJ: NTT Physics and Informatics Lab (F,E), MJ, MMF, and CL: (P)

\bmsection{Data availability} Data underlying the results presented in this paper are not publicly available at this time but may be obtained from the authors upon reasonable request.

\bmsection{Supplemental document}
See Supplement 1 for supporting content. 

\end{backmatter}

\bibliography{sample}



\end{document}